\documentstyle[12pt]{article}
\begin{document}

\title{Inflation, Supergravity and Superstrings}
\author{Ewan D. Stewart \\ Department of Physics \\
Kyoto University \\ Kyoto 606, Japan}
\maketitle
\begin{abstract}
The positive potential energy required for inflation spontaneously
breaks supersymmetry and in general gives any would-be inflaton
an effective mass of order the inflationary Hubble parameter thus
ruling it out as an inflaton. In this paper I give simple conditions
on the superpotential that eliminate some potential sources for this
mass, and derive a form for the K\"{a}hler potential that eliminates
the rest. This reduces the problem of constructing a model of inflation
in supergravity to that of constructing one in global supersymmetry
with the extra conditions $ W = W_\varphi = \psi = 0 $ during inflation
(where $W$ is the superpotential, the inflaton $\in \varphi$, and
$W_\psi \neq 0$). I then point out that K\"{a}hler potentials of the
required form often occur in superstrings and that the target space
duality symmetries of superstrings often contain $R$-parities which
would make $ W = W_\varphi = 0 $ automatic for $ \psi = 0 $.
\end{abstract}
\vspace*{-88ex}
\hspace*{\fill}{\bf KUNS 1261}\hspace*{1.7em}\\
\hspace*{\fill}{hep-ph/9405389}
\thispagestyle{empty}
\setcounter{page}{0}
\newpage
\setcounter{page}{1}

\section{Introduction}

The approximate isotropy of the cosmic microwave background radiation
implies that the inflation \cite{Lindebook,Kolbbook} that inflated the
observable universe beyond the Hubble radius must have occurred at an
energy scale $V^{1/4} \leq 4 \times 10^{16}\,$GeV \cite{Liddle}, and it
is thought that physics at energies below the Planck scale is described
by an effective $N=1$ supergravity theory \cite{susy}. Thus models of
inflation should be constructed in the context of supergravity.
However, this immediately leads to a problem. The positive potential
energy $ V > 0 $ required for inflation spontaneously breaks
supersymmetry,\footnote{After inflation $V$ disappears and so
supersymmetry is restored modulo whatever breaks supersymmetry in our
vacuum.} which would generally be expected to give effective masses
$\sim \sqrt{8\pi V} / m_{\rm Pl} \sim H$ to any would-be inflatons.
But inflation requires $|V''/V| \ll 1$, {\mbox i.e.} the effective mass
of the inflaton must be much less than the inflationary Hubble parameter
$H$.

Natural inflation \cite{natural} avoids this problem by assuming the
inflaton corresponds to an angular degree of freedom whose potential
is kept flat enough by an approximate compact global symmetry.
The model of Holman et al.\ \cite{Ross,Kolbbook} assumes the form of
supergravity that gives minimal kinetic terms\footnote{This is no longer
regarded as realistic.} and fine tunes a parameter in the superpotential
to eliminate the troublesome mass term. Solutions to this problem which
work for $\phi^n$ chaotic inflation have also been proposed
\cite{Linde,Lindebook,Yanagida}, but they rely on forms for the
supergravity K\"{a}hler potential that have no independent motivation.
In this paper I will propose a solution for inflaton fields which are not
purely angular degrees of freedom, which requires no fine tuning, and
which uses a well motivated form for the K\"{a}hler potential. Some
aspects of this solution have been investigated in \cite{fvi}.

\section{Basic Formulae and Notation}

I will use the following conventions in this paper:
$m_{\rm Pl}/\sqrt{8\pi} = 1$, a prime will denote the derivative with
respect to the canonically normalised inflaton field~$\sigma$, a bar
will denote the hermitian conjugate, $\phi$ will represent a vector
whose components $\phi^\alpha$ are complex scalar fields, and
subscript $\phi$ will denote the derivative with respect to $\phi$, so
for example $W_\phi$ represents the vector with components
$ \partial W / \partial \phi^\alpha $.

\subsection{Global supersymmetry}

In global supersymmetry \cite{susy} the scalar kinetic terms are
\begin{equation}
\left| \partial_\mu \phi \right|^2 \,,
\end{equation}
where $\phi = ( \phi^1 , \phi^2 , \ldots )$ and the $\phi^\alpha$ are
complex scalar fields.
The scalar potential is
\begin{equation}
\label{Vsusy}
V = \left| W_\phi \right|^2 + \frac{1}{2} \sum_a g_{a}^{2} D_{a}^{2}
\,,
\end{equation}
with
\begin{equation}
D_a = \bar{\phi} Q_a \phi + \xi_a \,,
\end{equation}
where the superpotential $W(\phi)$ is an analytic function of $\phi$,
$a$ labels the gauge group generators $Q_a$, $g_a$ is the gauge
coupling constant, and the real constant $\xi_a$ is a Fayet-Iliopoulos
term that can be non-zero only if $Q_a$ generates a U(1) gauge group.
The first term is called the $F$-term and the second the $D$-term.
I will assume that the $F$-term gives rise to the inflationary
potential energy density and that the $D$-term is flat along the
inflationary trajectory so that it can be ignored during inflation.
It may however play a vital role in determining the trajectory and
in stabilising the non-inflaton fields.

\subsection{Supergravity}

The scalar fields in a supergravity theory are the coordinates of a
K\"{a}hler manifold. The metric on a K\"{a}hler manifold is
$ K_{\bar{\phi} \phi} $ where the K\"{a}hler potential
$K(\phi,\bar{\phi})$ is a real function of $\phi$ and its
hermitian conjugate $\bar{\phi}$. The scalar kinetic terms are
\begin{equation}
\partial_\mu \bar{\phi} \, K_{\bar{\phi} \phi} \,
\partial^\mu \phi \,,
\end{equation}
the $F$-term part of the scalar potential is
\begin{equation}
\label{VF}
V_F = e^K \left[ \left( W_\phi + W K_\phi \right)
	K^{-1}_{\bar{\phi} \phi}
	\left( \bar{W}_{\bar{\phi}} + \bar{W} K_{\bar{\phi}} \right)
	- 3 |W|^2 \right] \,,
\end{equation}
and the $D$-term part is
\begin{equation}
V_D = \frac{1}{2} \sum_{a,b} \left( {\rm Re}\, f_{ab} \right)^{-1}
	D_a D_b \,,
\end{equation}
with
\begin{equation}
D_a = K_{\phi} Q_a \phi + \xi_a \,,
\end{equation}
where $f_{ab}(\phi)$ is an analytic function of $\phi$ transforming as
a symmetric product of two adjoint representations of the gauge group.
Only the combination
\begin{equation}
G(\phi,\bar{\phi}) = K + \ln |W|^2
\end{equation}
is physically relevant and we are always free to make a K\"{a}hler
transformation:
\begin{equation}
K(\phi,\bar{\phi}) \rightarrow K(\phi,\bar{\phi}) - F(\phi)
	- \bar{F}(\bar{\phi}) \,,\;\;\;\;
W(\phi) \rightarrow e^{F(\phi)} W(\phi) \,.
\end{equation}

\subsection{Inflation}
I assume the effective action during inflation
\cite{Lindebook,Kolbbook} to have the form
\begin{equation}
S = \int d^4 x \sqrt{-g} \left[ - \frac{1}{2} R + g^{\mu\nu}
	\partial_\mu \bar{\phi} \, K_{\bar{\phi} \phi} \,
	\partial_\nu \phi - V(\phi,\bar{\phi}) \right] \,,
\end{equation}
and make the usual flat Robertson-Walker ansatz
\begin{equation}
ds^2 = dt^2 - a(t)^2 d{\bf x}^2 \,,\;\;\;\; \phi = \phi (t) \,.
\end{equation}
The Hubble parameter $H$ is defined as $ H \equiv \dot{a} / a $.
Inflation requires\footnote{Strictly speaking
$ -\dot{H} / H^2 < 1 $ is all that is required. However realistic
models satisfy $ -\dot{H} / H^2 \ll 1 $. See \cite{dhl} for a more
detailed discussion.}
$ -\dot{H} / H^2 \ll 1 $, or equivalently $ 3 H^2 \simeq V $,
{\mbox i.e.} the energy density of the universe should be dominated
by the scalar potential. The dynamics of the scalar fields then
rapidly approaches the slow-roll equations of motion
\begin{equation}
\dot{\phi} = - 3 H K^{-1}_{\bar{\phi} \phi} V_{\bar{\phi}} \,,
\end{equation}
and I will assume that they have been attained for all epochs of
interest. The canonically normalised inflaton $\sigma$ is defined by
\begin{equation}
\frac{1}{2} d\sigma^2 = d\bar{\phi} \, K_{\bar{\phi} \phi} \, d\phi
\,.
\end{equation}
The conditions necessary for inflation can be expressed in terms of
the potential as
\begin{equation}
\left( \frac{V'}{V} \right)^2 \ll 1 \,,\;\;\;\;
\left| \frac{V''}{V} \right| \ll 1 \,,
\end{equation}
where a prime denotes the derivative with respect to $\sigma$.

\section{The Problem}
\label{prob}

At any point in the space of scalar fields we can make a holomorphic
field redefinition such that $\phi=0$ and the scalar fields have
canonical kinetic terms at that point. Any purely holomorphic terms in
the K\"{a}hler potential can then be absorbed into the superpotential
using a K\"{a}hler transformation. Then, in the neighbourhood of that
point, the K\"{a}hler potential will be
\begin{equation}
K = \left| \phi \right|^2 + \ldots \,,
\end{equation}
where \ldots\ stand for higher order terms. Therefore from
Eq.~(\ref{VF})
\begin{eqnarray}
V & = & \exp \left( \left| \phi \right|^2 + \ldots \right) \times \\
 & & \left\{ \left[ W_\phi + W \left( \bar{\phi} + \ldots \right)
	\right] \left( 1 + \ldots \right) \left[ \bar{W}_{\bar{\phi}}
	+ \bar{W} \left( \phi + \ldots \right) \right] - 3 |W|^2 \right\}
 \,, \nonumber \\
 & = & V|_{\phi=0} + V|_{\phi=0} \left| \phi \right|^2
	+ {\rm other\ terms} \,.
\end{eqnarray}
Thus at $\phi=0$ the exponential term gives a contribution $V$ to the
effective mass squared of {\em all} scalar fields. Therefore,
\begin{equation}
\frac{ V''}{V} = 1 + {\rm other\ terms} \,,
\end{equation}
where the prime denotes the derivative with respect to the canonically
normalised inflaton field. But $|V''/V| \ll 1$ is necessary for inflation
to work. So a sucessful model of inflation must arrange for a cancellation
between the exponential term and the terms inside the curly brackets.
This will require fine tuning unless a symmetry is used to enforce it.
Natural inflation \cite{natural} uses an approximate compact global
symmetry. I will use a combination of a discrete $R$-symmetry and a
non-compact global symmetry.

\section{A Solution}
\label{ps}

Divide the vector of scalar fields $\phi$ into two separate vectors,
$\varphi$ and $\psi$, with the inflaton contained in $\varphi$:
\begin{equation}
\phi = (\varphi,\psi) \,,\;\;\;\; {\rm inflaton} \in \varphi \,.
\end{equation}
Then, during inflation, $\psi$ is constant and so without loss of
generality we can set
\begin{equation}
\label{p0}
\psi = 0 \,.
\end{equation}
When we want to distinguish non-inflaton $\varphi$ fields from the
inflaton we will denote them by $\chi$:
\begin{equation}
\chi \subset \varphi \,,\;\;\;\; {\rm inflaton} \not\in \chi \,.
\end{equation}
Note that $\chi$ is constant during inflation.

Now the key point of the solution is to assume that
\begin{equation}
\label{ps1}
W = W_\varphi = 0 \;\;\;\;{\rm and}\;\;\;\; W_\psi \neq 0
\end{equation}
during inflation.
Then the scalar potential Eq.~(\ref{VF}) simplifies to
\begin{equation}
\label{Vi}
V = e^K W_\psi K^{-1}_{\bar{\psi} \psi} \bar{W}_{\bar{\psi}} \,.
\end{equation}
Now it becomes possible to choose a form for the K\"{a}hler
potential that cancels the inflaton dependent corrections to the
global supersymmetry potential in a natural way.

For simplicity, assume that
\begin{equation}
\label{ps2}
\left. K_{\bar{\varphi} \psi} \right|_{\psi=0} = 0 \,,
\end{equation}
so that $\varphi$ and $\psi$ have no mixed kinetic terms during
inflation.
Then, using a K\"{a}hler transformation to remove any remaining terms
linear in $\psi$, and expanding about $\psi = 0$, we get
\begin{equation}
\label{ps3}
K = A(\varphi,\bar{\varphi}) + \bar{\psi}\, B(\varphi,\bar{\varphi})
	\,\psi + {\cal O} \left( \psi^2 , \bar{\psi}^2 \right) \,,
\end{equation}
where $A$ is a real function and $B$ is a positive definite hermitian
matrix.

Note that Eqs.~(\ref{ps1}), (\ref{ps2}) and~(\ref{ps3}) become
automatic if we impose the symmetry (an $R$-parity)
\begin{equation}
\label{R}
\psi \rightarrow -\psi \,,\;\;\;\; \varphi \rightarrow \varphi
\,,\;\;\;\;W \rightarrow -W \,,\;\;\;\; K \rightarrow K \,,
\end{equation}
which also helps to stablise $\psi$ at 0 because $V_\psi = 0$ is
also automatic.\footnote{In fact, any unbroken discrete (or continuous)
$R$-symmetry of the form
$ W \rightarrow e^{i\theta_0} W $,
$ \psi \rightarrow e^{i\theta_0} \psi $,
$ \chi^\alpha \rightarrow e^{i\theta_\alpha} \chi^\alpha $,
$ \theta_\alpha \neq \theta_0 $
would suffice.}

{}From Eqs.~(\ref{Vi}) and~(\ref{ps3}),
\begin{equation}
V = e^A W_\psi B^{-1} \bar{W}_{\bar{\psi}} \,,
\end{equation}
and so to eliminate the inflaton dependent corrections to the global
supersymmetry potential we require
\begin{equation}
B^{-1} = f(\varphi,\bar{\varphi}) \, C^{-1}(\chi,\bar{\chi}) \,,
\end{equation}
and
\begin{equation}
A = - \ln f(\varphi,\bar{\varphi}) + g(\chi,\bar{\chi}) \,,
\end{equation}
where $f$ and $g$ are real functions, and $C$ is a positive definite
hermitian matrix. This gives the inflationary potential
\begin{equation}
V = e^{ g(\chi,\bar{\chi}) }
	W_\psi\, C^{-1}(\chi,\bar{\chi}) \,\bar{W}_{\bar{\psi}} \,,
\end{equation}
and the K\"{a}hler potential is required to have the general form
\begin{eqnarray}
K & = & - \ln f(\varphi,\bar{\varphi})
	+ \frac{ \bar{\psi}\, C(\chi,\bar{\chi}) \,\psi }
	{ f(\varphi,\bar{\varphi}) } + g(\chi,\bar{\chi})
	+ {\cal O} \left( \psi^2 , \bar{\psi}^2 \right) \,, \\
\label{K}
 & = & - \ln \left[ f(\varphi,\bar{\varphi})
	- \bar{\psi}\, C(\chi,\bar{\chi}) \,\psi \right]
+ g(\chi,\bar{\chi}) + {\cal O} \left( \psi^2 , \bar{\psi}^2 \right) \,.
\end{eqnarray}

\section{Simple Examples of Suitable K\"{a}hler Potentials}
\label{se}
\subsection{${\rm SU}(m,1) / \left({\rm SU}(m)\times{\rm U}(1)\right)$}
\label{su}

The simplest example of Eq.~(\ref{K}) is
\begin{equation}
\label{simpleK}
K = - \ln X \,,\;\;\;\; X = 1 - \left| \phi \right|^2 \,.
\end{equation}
The corresponding K\"{a}hler manifold is
$ {\rm SU}(m,1) / \left( {\rm SU}(m) \times {\rm U}(1) \right) $,
where $m$ is the number of components of $\phi$. It is a maximally
symmetric space with constant riemannian curvature. Such coset spaces
form the basis of no-scale supergravity \cite{noscale}, though it is
important to note that the K\"{a}hler potential in Eq.~(\ref{simpleK})
only corresponds to part of one sector of a no-scale model. Now
\begin{equation}
\label{kinv}
K_{\bar{\phi} \phi} = \frac{1}{X^2} \left( X + \phi \bar{\phi} \right)
	\,,\;\;\;\;
K^{-1}_{\bar{\phi} \phi} = X \left( 1 - \phi \bar{\phi} \right) \,,
\end{equation}
and from Eq.~(\ref{VF})
\begin{equation}
V = \left| W_\phi \right|^2 - \left| W_\phi \phi - W \right|^2
	- \frac{2}{X} \left| W \right|^2 \,.
\end{equation}
Let $\phi = (\varphi,\psi)$, and assume $W = W_\varphi = \psi = 0$.
Then the kinetic terms are
\begin{equation}
\label{kin}
\frac{1}{X^2} \, \partial_\mu \bar{\varphi}
\left( X + \varphi \bar{\varphi} \right) \partial^\mu \varphi \,,
\end{equation}
and the potential is
\begin{equation}
V =  \left| W_\psi \right|^2 \,.
\end{equation}
Thus for $|\varphi| \ll 1$ we have canonical kinetic terms and the
potential has the global supersymmetry form, though with the
additional requirements $W = W_\varphi = \psi = 0$.

\subsection{${\rm SO}(m,2) / \left({\rm SO}(m)\times{\rm SO}(2)\right)$}

Another example of Eq.~(\ref{K}) is
\begin{equation}
\label{ae}
K = - \ln \left( 1 - \sum_{\alpha=1}^{m} \phi^\alpha \bar{\phi}^\alpha
	+ \frac{1}{4} \left| \sum_{\alpha=1}^{m} \phi^\alpha \phi^\alpha
	\right|^2 \right) \,.
\end{equation}
The corresponding K\"{a}hler manifold is
$ {\rm SO}(m,2) / \left( {\rm SO}(m) \times {\rm SO}(2) \right) $. For
example, if $m=2$, $\varphi = \phi^1$ and $\psi = \phi^2$ we get
\begin{equation}
\label{so1}
K = - \ln \left[ \left( 1 - \frac{1}{2} |\varphi|^2 \right)^2
- |\psi|^2 \right] + {\cal O} \left( \psi^2 , \bar{\psi}^2 \right) \,,
\end{equation}
or if $m=3$,
$ \varphi = \left( \phi^1 + i\phi^2 \right) / \sqrt{2} $,
$ \chi = \left( \phi^1 - i\phi^2 \right) / \sqrt{2} $ and
$ \psi = \phi^3 $ we get
\begin{equation}
\label{so2}
K = - \ln \left[ \left( 1 - |\varphi|^2 \right)
	\left( 1 - |\chi|^2 \right) - |\psi|^2 \right]
	+ {\cal O} \left( \psi^2 , \bar{\psi}^2 \right) \,.
\end{equation}

\section{Model Building}

The solution described in Section~\ref{ps} suggests a natural
strategy for inflationary model building. Construct a globally
supersymmetric model which gives rise to inflation and satisfies
Eqs.~(\ref{p0}) and~(\ref{ps1}), at least to some approximation
- see Section~\ref{sip}. Then extend to supergravity by
choosing a K\"{a}hler potential of the form of Eq.~(\ref{K}).
However, as we shall see in Section~\ref{nmi}, it may not even
be necessary for the globally supersymmetric model to give rise
to inflation.

\subsection{An example of a suitable globally supersymmetric model}

Consider the following globally supersymmetric model
\begin{equation}
W = \lambda_1 \varphi \chi_1 \psi_1 + \lambda_2 \chi_{2}^{n} \psi_2
 \,,
\end{equation}
and
\begin{equation}
D = \Lambda^2 - \left| \chi_1 \right|^2 - \left| \chi_2 \right|^2
	+ \left| \psi_1 \right|^2 + n \left| \psi_2 \right|^2 \,.
\end{equation}
This model is invariant under the $R$-parity
\begin{equation}
\psi_1 \rightarrow -\psi_1 \,,\;\;\;\; \psi_2 \rightarrow -\psi_2
\,,\;\;\;\; W \rightarrow -W \,,
\end{equation}
and the U(1) gauge symmetry\footnote{For example an `anomalous' U(1)
often appears in string theory \cite{Dine} with
$ \Lambda \sim 10^{17}-10^{18}\,$GeV {\em if\/} the dilaton is fixed
near its usual value.}
\begin{equation}
\chi_1 \rightarrow e^{-i\theta} \chi_1 \,,\;\;\;\;
\chi_2 \rightarrow e^{-i\theta} \chi_2 \,,\;\;\;\;
\psi_1 \rightarrow e^{i\theta} \psi_1 \,,\;\;\;\;
\psi_2 \rightarrow e^{in\theta} \psi_2 \,.
\end{equation}

To obtain the effective potential during inflation we minimise the
potential [Eq.~(\ref{Vsusy})] for fixed $\varphi$ as follows.
For $\left| \chi_1 \right|^2 + \left| \chi_2 \right|^2 \leq \Lambda^2 $
the potential is minimised for
\begin{equation}
\label{a1}
\psi_1 = \psi_2 = 0 \,,
\end{equation}
and so the $R$-parity ensures that
\begin{equation}
\label{a2}
W = W_{\varphi} = W_{\chi_1} = W_{\chi_2} = 0 \,.
\end{equation}
Therefore
\begin{eqnarray}
V & = & \left| W_{\psi_{1}} \right|^2 + \left| W_{\psi_{2}} \right|^2
	+ \frac{1}{2} g^2 D^2 \,, \\
 & = & \lambda_{1}^{2} \left| \varphi \right|^2 \left| \chi_1
	\right|^2 + \lambda_{2}^{2} \left| \chi_2 \right|^{2n}
+ \frac{1}{2} g^2 \left( \Lambda^2 - \left| \chi_1 \right|^2
	- \left| \chi_2 \right|^2 \right)^2  \,.
\end{eqnarray}
Now if
\begin{equation}
\label{pgl}
|\varphi|^2 \geq \frac{ g^2 }{ \lambda_{1}^{2} }
	\left( \Lambda^2 - \left| \chi_2 \right|^2 \right) \,,
\end{equation}
the potential is minimised for
\begin{equation}
\chi_1 = 0 \,.
\end{equation}
Then
\begin{equation}
V = \lambda_{2}^{2} \left| \chi_2 \right|^{2n} + \frac{1}{2} g^2
	\left( \Lambda^2 - \left| \chi_2 \right|^2 \right)^2 \,.
\end{equation}
Now if
\begin{equation}
\label{lcon}
\frac{ \lambda_2 \Lambda^{n-2} }{ g } \ll 1 \,,
\end{equation}
the potential is minimised for
\begin{equation}
\label{chi}
\left| \chi_2 \right|^2 \simeq \Lambda^2
	- \frac{ n \lambda_{2}^{2} \Lambda^{2n-2} }{ g^2 } \,,
\end{equation}
and so
\begin{equation}
V \simeq \lambda_{2}^{2} \Lambda^{2n} \,.
\end{equation}
Thus, from Eqs.~(\ref{pgl}) and~(\ref{chi}), for
\begin{equation}
|\varphi| \geq
\frac{ \sqrt{n}\, \lambda_2 \Lambda^{n-1} }{ \lambda_1 } \,,
\end{equation}
we have a positive potential energy density and a flat potential
for the inflaton field $\varphi$.

The above globally supersymmetric model satisfies the conditions
Eqs.~(\ref{p0}) and~(\ref{ps1}) [Eqs.~(\ref{a1}) and~(\ref{a2})] and
so if the K\"{a}hler potential is of the form of Eq.~(\ref{K}) then
the supergravity corrections will not spoil the flatness of the
inflaton's potential (which is exactly flat in this case but there are
many possible sources for a small slope for the inflaton's potential
- see the next section). Also, it is easy to check that the
supergravity corrections do not spoil the stability of the model.

Alternatively to Eq.~(\ref{lcon}), if $n=1$ and
$ g \Lambda / \lambda_2 < 1 $ then the potential is minimised for
$\chi_2 = 0$ and
\begin{equation}
V = \frac{1}{2} g^2 \Lambda^4 \,.
\end{equation}
In this case the inflationary potential energy density is dominated by
the $D$-term part of the scalar potential which might provide an
alternative solution to the problem discussed in Section~\ref{prob}.

\subsection{The slope of the inflaton's potential}
\label{sip}

The solution described in Section~\ref{ps} is unlikely to hold exactly
in realistic models. Small deviations from it lead to small
contributions to the slope of the inflaton's potential. In some cases
these could dominate the contributions coming from the globally
supersymmetric model and so effectively determine the slope of the
inflaton's potential. For example, if $ W = W_\varphi = \psi = 0 $
but $ K = K_0 + \delta(\varphi,\bar{\varphi}) $ where $K_0$ is of the
form of Eq.~(\ref{K}), then we get $ V = e^\delta V_0 $ where $ V_0 $ is
the potential which would have been obtained if $ K = K_0 $ had been used.
We thus get a contribution of $\delta'$ to $V'/V$ and of $\delta''$
to $ V''/V - (V'/V)^2 $. See \cite{fvi} for an explicit example.
Also, $ W \neq 0 $ or $ W_\varphi \neq 0 $ would typically give a
contribution to $V''/V$ of order $ |W|^2 / |W_\psi |^2 $ or
$ | W_\varphi |^2 / | W_\psi |^2 $ respectively.

\subsection{Inflation without inflation in the global supersymmetry limit}
\label{nmi}

Another example of a globally supersymmetric model satisfying
Eqs.~(\ref{p0}) and~(\ref{ps1}) is
\begin{equation}
W = \lambda \, f(\varphi) \, \chi^n \psi \,,
\end{equation}
and
\begin{equation}
D = \Lambda^2 - \left| \chi \right|^2 + n \left| \psi \right|^2 \,.
\end{equation}
For $ \lambda \Lambda^{n-2} |f(\varphi)| / g \ll 1 $ it gives a potential
\begin{equation}
V \simeq \lambda^2 \Lambda^{2n} \left| f(\varphi) \right|^2 \,.
\end{equation}
This will not give rise to inflation in the global supersymmetry
limit\footnote{which requires $ |\varphi| \ll 1 $ for consistency}
for a generic function $f(\varphi)$.
However, in the supergravity theory with the K\"{a}hler potential discussed
in Section~\ref{su}, the kinetic terms are non-canonical and diverge as
$|\varphi|$ approaches one,\footnote{Note that $\varphi$ is defined only
for $|\varphi| < 1$.}
but the $\varphi$ dependence of the potential is unchanged from the global
supersymmetry limit. Therefore, transforming to the canonically normalised
inflaton field stretches out the potential and so, assuming that
$f(\varphi)$ does not diverge as $|\varphi| \rightarrow 1$, we will get a
flat potential.

To illustrate this, consider the following simple example.
For the case of only one $\varphi$ field Eq.~(\ref{kin}) reduces to
\begin{equation}
\frac{1}{X^2} \left| \partial \varphi \right|^2 \,.
\end{equation}
For simplicity assume the phase of $\varphi$ is constant. Then
the canonically normalised inflaton $\sigma$ is given by
\begin{equation}
|\varphi| = \tanh \frac{\sigma}{\sqrt{2}} \,.
\end{equation}
Now during inflation $\sigma \gg 1$ and so
\begin{equation}
|\varphi| \simeq 1 - 2 e^{-\sqrt{2}\,\sigma} \,.
\end{equation}
Therefore
\begin{equation}
V \simeq V|_{|\varphi|=1} - 2 \left.
\frac{d V}{d |\varphi|} \right|_{|\varphi|=1} e^{-\sqrt{2}\,\sigma} \,.
\end{equation}
The coefficient of the exponential can be absorbed by the redefinition
\begin{equation}
\tilde{\sigma} = \sigma - \frac{1}{\sqrt{2}} \ln \left.
\frac{ 2 \frac{d V}{d |\varphi|} }{ V } \right|_{|\varphi|=1} \,,
\end{equation}
to give the inflationary potential\footnote{Note that this is
the potential during inflation ($\tilde{\sigma} \gg 1$). When
inflation ends ($\tilde{\sigma} \sim 1$), the neglected, model
dependent ({\mbox i.e.} $f$ dependent) terms become important.}
\begin{equation}
V = V_1 \left( 1 - e^{ -\sqrt{2}\, \tilde{\sigma} } \right) \,,
\end{equation}
which has only one free parameter $V_1 = V|_{|\varphi|=1}$ and that is
determined by the COBE normalisation to be
$ V_{1}^{1/4} = 6 \times 10^{15} \,$GeV.
It is also straightforward to calculate the spectral index of the
density perturbations produced during inflation \cite{n,dhl},
\begin{equation}
n = 1 - 3 \left( \frac{V'}{V} \right)^2 + 2 \frac{V''}{V}
\simeq 1 - \frac{2}{N} \simeq 0.96 \,,
\end{equation}
which is the same as $\phi^2$ chaotic inflation. However, the ratio
of gravitational waves to density perturbations is \cite{n,dhl}
\begin{equation}
R = 6 \left( \frac{V'}{V} \right)^2 = \frac{3}{N^2} \sim 10^{-3} \,,
\end{equation}
compared with $R = 6/N = 0.1$ for $\phi^2$ chaotic inflation.

It is interesting that these results are quite robust, at least for a
single inflationary degree of freedom. For example, if we had instead
chosen the K\"{a}hler potential of Eq.~(\ref{so1}), we would have got
the inflationary potential $ V = V_1 \left( 1 - e^{-\sigma} \right) $
which also gives $ n = 1 - 2/N \simeq 0.96 $ but slightly larger
$ V_{1}^{1/4} = 7 \times 10^{15} \, $GeV and
$ R = 6/N^2 \sim 10^{-2.5} $.

\section{Superstring Examples}

\subsection{Orbifold compactifications}
\label{oc}

The K\"{a}hler potential of the untwisted sector of the low-energy
effective supergravity theory derived from orbifold compactification
of superstrings always contains \cite{Ferrara}
\begin{equation}
K = - \ln \left( S + \bar{S} \right)
- \sum_{i=1}^{3} \ln \left( T_i + \bar{T}_i
- \left| \phi_i \right|^2 \right) \,,
\end{equation}
where $S$ is the dilaton, $T_i$ are the untwisted moduli associated
with the radii of compactification, and $\phi_i$ are the untwisted
matter fields associated with $T_i$.
Now if we divide the scalar fields into $\varphi$, $\psi$ and $\chi$
fields as follows
\begin{eqnarray}
T_1 & \in & \varphi \,, \\
\phi_1 & \in & \psi \,, \\
S \,,\, T_2 \,,\, T_3 \,,\, \phi_2 \;{\rm and}\; \phi_3 & \in &
\chi \subset \varphi \,,
\end{eqnarray}
then we get a K\"{a}hler potential of the required form
[Eq.~(\ref{K})]
\begin{equation}
K = - \ln \left( \varphi + \bar{\varphi} - \left| \psi \right|^2
	\right) + g(\chi,\bar{\chi}) \,,
\end{equation}
and the target space duality symmetries \cite{duality},
\begin{equation}
T_i \rightarrow \frac{a_i T_i - i b_i}{i c_i T_i + d_i} \,,\;\;\;\;
\phi_i \rightarrow \frac{\phi_i}{i c_i T_i + d_i} \,,\;\;\;\;
a_i d_i - b_i c_i = 1 \,,
\end{equation}
contain the desired $R$-parity [Eq.~(\ref{R})] on setting
$ b_i = c_i = 0 $, $ a_1 = d_1 = -1 $ and
$ a_2 = a_3 = d_2 = d_3 = 1 $.

\subsection{More orbifold compactifications}

A K\"{a}hler potential of the form
\begin{equation}
K = - \ln \left[ \left( T + \bar{T} \right) \left( U + \bar{U} \right)
	- \left( B + \bar{C} \right) \left( C + \bar{B} \right) \right]
	\,,
\end{equation}
often occurs in orbifold compactifications \cite{Ferrara,Cvetic,Cardoso}.
In particular, it can arise in orbifolds with continuous Wilson lines,
in which case $T$ corresponds to one of the $T_i$ moduli of
Section~\ref{oc}, $U$ is a (1,2) modulus, and $B$ and $C$ are continuous
Wilson line moduli \cite{Cardoso}.
Now if we divide the fields as follows
\begin{eqnarray}
T \;{\rm and}\; U & \in & \varphi \,, \\
B \;{\rm and}\; C & \in & \psi \,,
\end{eqnarray}
then we get a K\"{a}hler potential of the required form [Eq.~(\ref{K})]
\begin{equation}
K = - \ln \left[ \left( \varphi^1 + \bar{\varphi}^1 \right)
	\left( \varphi^2 + \bar{\varphi}^2 \right)
	- \left| \psi \right|^2 \right]
+ {\cal O} \left( \psi^2 , \bar{\psi}^2 \right) \,,
\end{equation}
and the target space duality symmetries \cite{Cardoso} contain the
desired $R$-parity [Eq.~(\ref{R})].

\subsection{Fermionic four-dimensional string models}

The K\"{a}hler potential of the untwisted sector of the revamped
flipped SU(5) model \cite{flipped} is \cite{Lopez}
\begin{eqnarray}
\lefteqn{ K = -\ln \left( 1 - \left| \Phi_1 \right|^2
- \left| \Phi_{23} \right|^2 - \left| \Phi_{\overline{23}} \right|^2
- \left| h_1 \right|^2 - \left| h_{\overline{1}} \right|^2
+ \frac{1}{4} \left| \Phi_{1}^{2} + 2 \Phi_{23} \Phi_{\overline{23}}
	+ 2 h_1 h_{\overline{1}} \right|^2 \right) } \nonumber \\
 & & \mbox{} - \ln \left( 1 - \left| \Phi_2 \right|^2
- \left| \Phi_{31} \right|^2 - \left| \Phi_{\overline{31}} \right|^2
- \left| h_2 \right|^2 - \left| h_{\overline{2}} \right|^2
+ \frac{1}{4} \left| \Phi_{2}^{2} + 2 \Phi_{31} \Phi_{\overline{31}}
	+ 2 h_2 h_{\overline{2}} \right|^2 \right) \nonumber \\
 & & \mbox{} - \ln \left( 1 - \left| \Phi_4 \right|^2
- \left| \Phi_5 \right|^2 - \left| \Phi_3 \right|^2
- \left| \Phi_{12} \right|^2 - \left| \Phi_{\overline{12}} \right|^2
- \left| h_3 \right|^2 - \left| h_{\overline{3}} \right|^2
\right. \nonumber \\
 & & \left. \hspace{3em} \mbox{} + \frac{1}{4} \left| \Phi_{4}^{2}
	+ \Phi_{5}^{2} + \Phi_{3}^{2} + 2 \Phi_{12} \Phi_{\overline{12}}
	+ 2 h_3 h_{\overline{3}} \right|^2 \right) \,,
\end{eqnarray}
all three parts of which are of the form of Eq.~(\ref{ae}).\footnote{
up to trivial redefinitions} Furthermore, if we divide the fields as follows
\begin{eqnarray}
\Phi_4 \;{\rm and}\; \Phi_5 & \in & \varphi \,, \\
\Phi_3 \,,\, \Phi_{12} \,,\, \Phi_{\overline{12}} \,,\, h_3
	\;{\rm and}\; h_{\overline{3}} & \in & \psi \,, \\
\Phi_1 \,,\, \Phi_2 \,,\, \Phi_{23} \,,\, \Phi_{\overline{23}} \,,\,
	\Phi_{31} \,,\, \Phi_{\overline{31}} \,,\, h_1 \,,\,
	h_{\overline{1}} \,,\, h_2 \;{\rm and}\; h_{\overline{2}} & \in &
	\chi \subset \varphi \,,
\end{eqnarray}
then we get a K\"{a}hler potential of the required form [Eq.~(\ref{K})]
\begin{equation}
K = - \ln \left( 1 - \left| \varphi \right|^2
	+ \frac{1}{4} \left| \varphi^{\rm T} \varphi \right|^2
	- \left| \psi \right|^2 \right) + g(\chi,\bar{\chi})
+ {\cal O} \left( \psi^2 , \bar{\psi}^2 \right) \,,
\end{equation}
and the target space duality symmetries \cite{Lopez} contain the
desired $R$-parity [Eq.~(\ref{R})].

\subsection{Calabi-Yau compactifications}

Here I give the Calabi-Yau manifold discussed in Section~4.3 of
\cite{Dixon} as another example of a compactification of superstrings
that can give K\"{a}hler potentials of the form discussed in
Section~\ref{se}. At a particular point in the moduli space of the
above mentioned Calabi-Yau manifold, the low-energy gauge group
includes an extra U$(1)^4$ factor, a subgroup of which may be
preserved on subspaces of the moduli space that pass through that
point. The K\"{a}hler potential on a subspace of the moduli space that
preserves a U$(1)^3$ subgroup of the extra U$(1)^4$ gauge symmetry is
\cite{Dixon}
\begin{equation}
K = - \ln \left( 1 - |N|^2 - |C|^2 \right)
	+ {\cal O} \left( C^2 , \bar{C}^2 \right) \,,
\end{equation}
where $ N = ( N_1 , N_2 )$ are the neutral\footnote{with respect to
the unbroken U$(1)^3$} (1,2) moduli that span the subspace, and $C$ is
a vector whose components are 63 of the 99 charged (1,2) moduli and
their associated matter fields.
Also the K\"{a}hler potential on a subspace of the moduli space that
preserves a U$(1)^2$ subgroup of the extra U$(1)^4$ gauge symmetry is
\cite{Dixon}
\begin{equation}
K = - \ln \left( 1 - |N|^2 \right) \,,
\end{equation}
where $N$ is a vector whose components are the twelve neutral (1,2)
moduli that span the subspace.

\section{Conclusions}

A globally supersymmetric model of inflation (see for example
\cite{fvi,sneu}) will not work in a generic supergravity theory
because the higher order, non-renormalisable supergravity corrections
destroy the flatness of the inflaton's potential. In this paper I have
derived a form for the K\"{a}hler potential which eliminates these
corrections if $ W = W_\varphi = \psi = 0 $ during inflation (where $W$
is the superpotential, the inflaton $\in \varphi$, and $W_\psi \neq 0$).
It is encouraging that K\"{a}hler potentials of the required form often
occur in superstrings and that the target space duality symmetries of
superstrings often contain $R$-parities which would make
$ W = W_\varphi = 0 $ automatic for $ \psi = 0 $.

Also, I have shown that supergravity theories with K\"{a}hler potentials
of this form may give rise to inflation even if the corresponding
globally supersymmetric theory does not. The simplest examples of this
new idea for inflation give a spectral index $ n = 1 - 2/N \simeq 0.96 $
for the density perturbations and negligible gravitational waves, though
more complicated examples lose this predictive power.

\subsection*{Acknowledgements}
I thank D. H. Lyth for detailed comments on various drafts of this paper,
and K. Maeda and M. Sasaki for helpful comments on an early draft of this
paper. I also thank the third referee for helpful comments.
I am supported by a JSPS Postdoctoral Fellowship and this work was
supported by Monbusho Grant-in-Aid for Encouragement of Young
Scientists No.\ 92062.

\end{document}